# A topological formal treatment for scenario-based software specification of concurrent real-time systems

**Miriam C. Bergue Alves**
**Christine C. Dantas**
**Nanci Naomi Arai**
**Rovedy B. da Silva**
Institute of Aeronautics and Space - IAE
Pca. Mal. Eduardo Gomes, 50 Vila das Acacias
12.228-904 São Jose dos Campos, Brazil
Tel: +55 12 39474969 –Fax: +55 12 39475019
e-mail: {miriamalves;ccdantas;naomi;rovedy}@iae.cta.br

**Abstract:** Real-time systems are computing systems in which the meeting of their requirements is vital for their correctness. Consequently, if the real-time requirements of these systems are poorly understood and verified, the results can be disastrous and lead to irremediable project failures at the early phases of development. The present work addresses the problem of detecting deadlock situations early in the requirements specification phase of a concurrent real time system, proposing a simple "proof-of-concepts" prototype that joins scenario-based requirements specifications and techniques based on topology. The efforts are concentrated in the integration of the formal representation of Message Sequence Chart scenarios into the deadlock detection algorithm of Fajstrup et al. [15], based on geometric and algebraic topology.

## 1. INTRODUCTION

A predominant characteristic of real-time systems is concurrency. Concurrent systems are composed of concurrent tasks, processes or objects, and they typically manage shared resources, which means a demand for predictability, flexibility and reliability [10] [5]. For the class of hard-real-time systems addressed by this research, there should be mechanisms and policies that ensure consistency and minimize worst case blocking, any boundless or excessive run-time overheads. Since such aspects are strongly associated with the behavioral model of the system, techniques and tools that more appropriately express concurrency, distribution and parallelism are indispensable.

The comprehension of concurrent systems is more difficult than the sequential ones for various reasons. Perhaps the most obvious is that in a concurrent system all the different components are in independent states and the combinations of states grow exponentially. Deadlocks can rise in these systems, which means that no component can make any progress, generally because each is waiting for communication with others. The formal understanding and reasoning can help to establish properties of such a system. A formal specification of a system is concerned with producing an ambiguous set of system specifications that can be formally verified, so that requirements, as well as the environment constraints and design intentions, are correctly reflected, thus reducing the chances of accidental fault injections. Recently, techniques based on algebraic topology [4] have been introduced into concurrency theory in order to deal with the high level complexity of verifying and analyzing properties of concurrent real-time systems. There is an entire assemblage of well-studied topological techniques that could be used, with some adaptations, to formally prove properties of concurrent systems. However, given the very recent acknowledgement and interest of computer scientists in topological techniques, there have been very few actual implementations of such ideas in real case situations.

The use of formal specification is an important feature that adds a great value at the initial stage of the software development. However, a certain level of non-formalism is necessary in the beginning of the development when the requirements are not completely understood and some flexibility is essential for trying out some alternatives to the design. Scenario-based requirements specification, expressed using Message Sequence Charts (MSCs), has a visual and proper semantics and has increasingly being used by analysts for specifying requirements of a software system [20] [8] [12] [2] [16]. MSCs have been major topic of research and practice [18] [21] and their semantics is also formalized [7] [6].

This work presents a simple method that maps scenario-based specifications, represented formally by MSCs, to a topological space in order to formally verify these specifications. The efforts are concentrated in the integration of the deadlock detection algorithm of Fajstrup et al. [15], based on topological techniques, to MSC scenarios, addressing the problem of detecting deadlock situations early in the requirements specification phase, proposing a simple "proof-of-concepts" prototype.

This paper is organized in six sections. Section 1 is the introduction, followed by Section 2 that presents the approach of modeling possible deadlocks scenarios using MSCs and process algebra. Section 3 establishes the necessary topological concepts for this work and Section 4 describes the integration between the deadlock detection algorithm and the MSCs models as a simple "proof-of-concepts" prototype. Finally, the conclusion and future prospects are summarized in Section 5.

## 2. MODELING POSSIBLE DEADLOCKS SCENARIOS WITH MSC AND PROCESS ALGEBRA

### 2.1 MSC as a graphical language for describing scenarios

Message Sequence Chart is one of the most widespread approaches to documenting scenario-based specifications, is relatively easy to use, has a wide acceptance in industry and is well suited for developing first approximations of intended behavior of a system. Scenarios describe a sequence of events or activities [11] [13] [9] and they refer to interactions between independent entities. A complete reference about the MSC language can be found in Recommendation Z120 [7].

### 2.2 The formalization of MSCs in process algebra

The ITU (International Telecommunication Union) Recommendation Z 120 [7] is the standardization for the MSC language. Due to its widespread use and popularity, the MSC semantics and static requirements were also formalized [6] [17]. [19]. The principal motivation behind this formalization is to offer a proper base for the language users to avoid ambiguities, inconsistencies and obscurities.

The description of the semantics of a MSC uses process algebra, based on the algebraic theory of process description ACP (Algebra of Communicating Process) [14]. A MSC is characterized by the sequence of events along an instance axis and it is assumed that there is an asynchronous communication between its instances.

In order to present some aspects of the MSC formalization related to this work, process algebra theory will be avoided and the necessary concepts will be introduced through some simple examples, using the MSC graphical representation. These examples do not exhaustively show all the elements of the basic language, but only presents the essential idea behind the formal semantics expressed in a process algebra sentence. An excellent tutorial about the formalization of MSC can be found in [19].

The MSC M1 in the Figure 1 describes two instances $P_1$ and $P_2$, which have two communications with the environment and one communication between them. This MSC can be characterized by two traces generated by $P_1$ and $P_2$. In process algebra, the semantics of $P_1$ and $P_2$ is:

Instance $P_1$: $out\ (P_1, P_2, m_2).\ in\ (env, P_1, m_1)$

Instance $P_2$: $in\ (P_1, P_2, m_2).\ in\ (env, P_2, m_3)$ where the operator "**.**" is strict sequential composition.

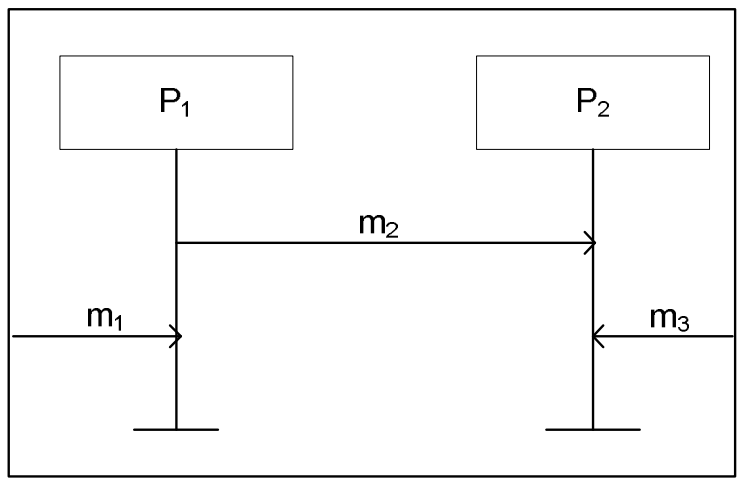

Figure 1: MSC M1

As $P_1$ and $P_2$ operate in parallel independently of each other, the semantics of the MSC M1 is:

$$out\ (P_1, P_2, m_2).\ in\ (env, P_1, m_1) \parallel in\ (P_1, P_2, m_2).\ in\ (env, P_2, m_3) \qquad (1)$$

where the operator "$\parallel$" is parallel composition.

The operator parallel defines an interleaved execution of its operands. There is a basic static requirement, which establishes that a message must be sent before it is received. Therefore, the expression (1), after expansion, has several traces that must be eliminated. In order to enforce this basic static requirement, the operator λ (state operator) is introduced. After applying λ, the semantics of the MSC M1 is established as follows:

$$out\ (P_1, P_2, m_2).[\ in\ (P_1, P_2, m_2)\ (\ in\ (env, P_1, m_1).\ in\ (env, P_2, m_3) + in\ (env, P_2, m_3).in\ (env, P_1, m_1))$$
$$+$$
$$in\ (env, P_1, m_1).in\ (P_1, P_2, m_2).\ in\ (env, P_2, m_3)]$$

where the operator "+" represents alternatives.

The MSC M2 in the Figure 2 describes two instances $P_1$ and $P_2$. As the instance $P_2$ has a coregion, the ordering of the events is completely free. Instead of using the sequential composition operator, the $\parallel$ operator is used.

The semantics of P1 is: $out\ (P_1, P_2, m_1).in\ (P_2, P_1, m_2).action\ (P_1, a)$. The semantics of $P_2$ is: $in\ (P_1, P_2, m_1) \parallel out\ (P_2, P_1, m_2)$. So, the semantics of the MSC M2 is:

$$out\ (P_1, P_2, m_1).[(in\ (P_1, P_2, m_1).\ out\ (P_2, P_1, m_2).\ in\ (P_2, P_1, m_2) +$$
$$out\ (P_2, P_1, m_2).\ in\ (P_1, P_2, m_1).\ in\ (P_2, P_1, m_2) +$$
$$out\ (P_2, P_1, m_2).\ in\ (P_2, P_1, m_2).\ in\ (P_1, P_2, m_1)].\ action\ (P_1, a)$$
$$+$$
$$out\ (P_2, P_1, m_2).[out\ (P_1, P_2, m_1).(in\ (P_1, P_2, m_1).\ in\ (P_2, P_1, m_2)$$
$$+\ in\ (P_2, P_1, m_2).\ in\ (P_1, P_2, m_1)].\ action\ (P_1, a)$$

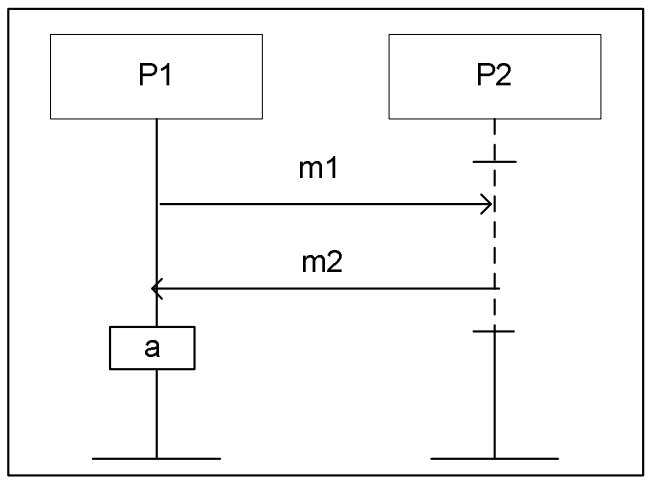

Figure 2: MSC M2

In general, the semantics of a MSC is a set of alternative traces that can be represented as shown in the Figure 3.

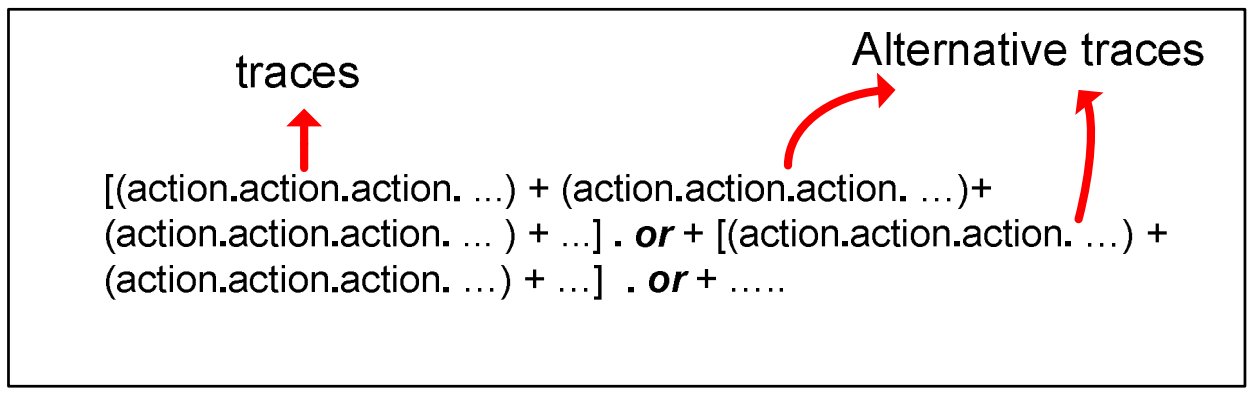

Figure 3: A general representation for a process algebra sentence.

## 2.3 Description of possible deadlock scenarios

There are two possible ways that concurrency can be raised from MSC scenarios: internal (within the MSC) and external (among different MSC scenarios). If deadlock conditions are detected early, changes in the system model can be made not only to eliminate them, but also to circumvent them in the future.

Some considerations have to be done in order to represent possible deadlock scenarios using MSCs. The identification of processes, resources and messages must reflect situations where there are a certain number of processes sharing resources in a mutual exclusion regime.

The processes will send two types of messages to the resources: **lock** and **unlock** (release). When a process sends a message **lock** to a resource, it will take that resource exclusively for a certain time in order to realize some processing and after that, this same process will release the resource by sending the message **unlock.** The **resources** are passive instances that only receive input messages for locking and unlocking themselves. As a result of these considerations, each instance of a MSC will to be a process or a resource. Figure 4 shows a scenario that illustrates two processes P1 e P2 sharing a resource R1.

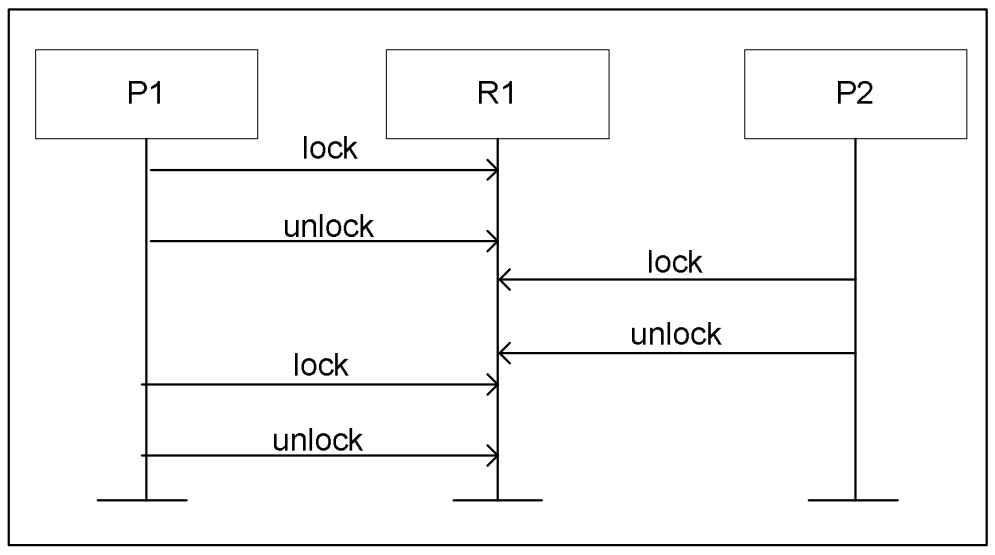

Figure 4: Lock-unlock scenario

### 2.4 The process algebra expression for possible deadlock scenarios

According to the assumptions established in the previous section, a MSC that represents a concurrent scenario will have only two possible types of messages: *lock* and *unlock*. So, the general expressions in process algebra can be established as follows:

Messages sent by processes*: out (process, resource, lock/unlock)*;
Messages received by resources*: in (process, resource, lock/unlock)*;

As the resources are considered passive instances, where the ordering that they will receive the messages **lock** and **unlock** is not determined in concurrent systems, they will be represented with coregion. So, assuming that there will be no lost messages, all the possible traces that correspond to all the ways the resource can be locked and unlocked will be the parallel composition of its process instances. This assumption eliminates the necessity of applying the operator λ.

In addition, the fact that the resource instances will receive all the messages sent to them in any order, and considering the geometric and topological treatment that will be used later (section 3 and 4), there will be no need to expand the whole process algebra expression for the semantics of the MSC. The semantics of its processes, written in a process algebra expression, will have all the information necessary to seek for deadlock scenarios.

There is a significant simplification when considering that the semantics of a MSC representing a possible deadlock scenario can be only characterized by the semantics of its identified processes. The resulting process algebra expression is now simple to understand and easier to create. The Figure 5 shows an example of a MSC that represents a possible deadlock scenario and the corresponding process algebra expression of its processes.

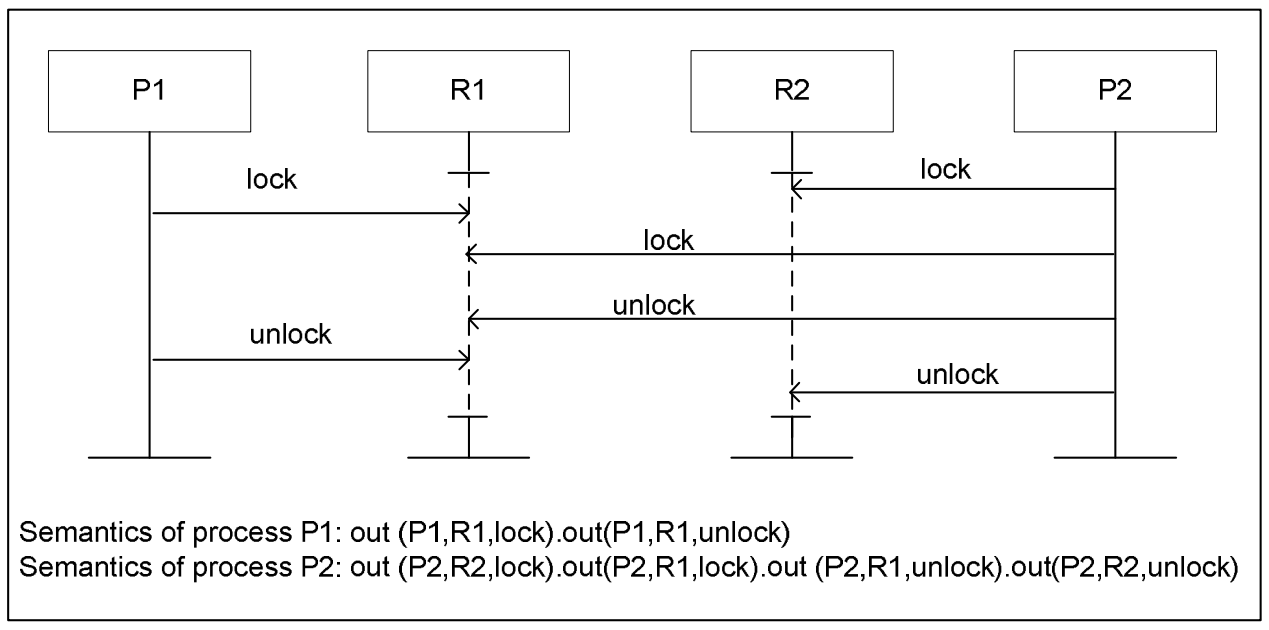

Figure 5: MSC and its algebra process sentence.

## 3. TOPOLOGICAL FORMAL TREATMENT

In recent years, topological methods have been introduced into concurrency theory (e.g., [4]). Most notably, the development of partial order reduction techniques based on topology have been used to tackle the well known "state-space explosion problem" [3]. Concurrency theory deals with a very large, although finite, space of states (a discrete space), whereas topology deals with the properties of geometrical figures that are preserved under continuous deformations. In order to apply the continuous topological techniques to the discrete space of concurrency, the latter is represented as a subset of the Euclidean space $R^n$: the unit cube in n-space, $I^n = I^1 x \dots x I^n$, where I is the unit interval [0,..,1]. Each coordinate axis corresponds to a process (from a set of n concurrent processes defined by the system). The set of coordinate points on each axis compose an ordered sequence of real numbers between 0 and 1, representing the scheduled actions (a transaction) that a given process will execute.

For instance, consider a finite set of transactions acted upon a centralized database. Each transaction can be abstracted as a sequence of locking (represented by P, according to Dijkstra's nomenclature [1]) and unlocking (V) to the database's shared resources (e.g., data records). The state of the database corresponds to a point in the n-cube space; particularly, the initial state is the n-dimensional vector with coordinates (0,…,0), whereas the final state is the (1,…,1) vector. If we consider that only two transactions, T1={PaPbVbVa} and T2={PbPaVaVb} (where a and b label the shared resources that are being locked and unlocked), will be acted upon the database, it is already clear from the geometrical representation that there will be three types of critical regions: unsafe, forbidden, and unreachable. An illustration of this two-transaction example is shown in Figure 6 (adapted from [15]).

Hence, if a concurrent system, composed of a certain number of independent processes, share one or several resources, a trajectory or path in the n-cube space corresponds to the correct synchronization between the processes only if the path does not cross the critical regions mentioned above. The forbidden region is actually a "hole" in the n-cube space that is inaccessible to the processes due to mutual exclusion. The unsafe region indicates a deadlock, whereas the unreachable region represents the set of impossible states of the system. Such a geometrical model of concurrency is referred to as a "progress graph".

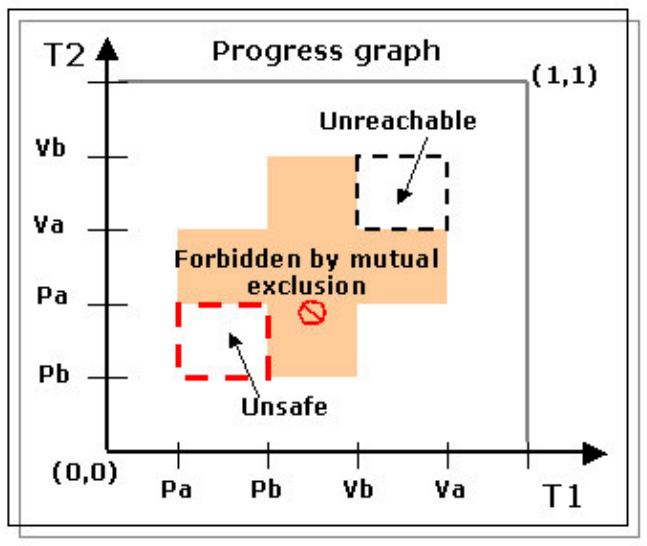

Figure 6: A two-transaction progress graph.

Given the existence of critical regions in the progress graph, it is already intuitively clear that it is possible to identify different sets of equivalent paths, in the sense that they will be performing essentially the same scheduling. For instance, if two or more execution paths can be continuously deformed into each other, then in topological terms they are homotopically equivalent. If a path cannot be deformed into another one, due to the presence of the excluded region between it and the other path, then it performs a different scheduling.

A progress graph is actually a topological space in which points representing the states of the concurrent system are ordered globally through time. Thus, it is already qualitatively evident from simple examples that the "state-space explosion problem" can be naturally tackled by such a topological formalism [3], given that there is no need to traverse all execution paths to check for given properties of the system. In particular, the existence of deadlocks can be geometrically determined by a simple algorithm developed by Fajstrup et al. [15].

## 4. INTEGRATING A DEADLOCK DETECTION ALGORITHM BASED ON TOPOLOGICAL METHODS TO MESSAGE SEQUENCE CHARTS

The main philosophy behind the "proof-of-concepts" prototype proposed in this paper is the formal verification of the implementability of MSC specifications, concerning to deadlocks scenarios, at an early stage of development of a concurrent system, based on ready-to-use topological concepts.

Such an integrability can be achieved in a relatively straightforward manner by recognizing that the fundamental actions of the semantics of each process identified in the MSC scenario (section 2.3) will be mapped in one coordinate axis, which corresponds to a process from a set of n concurrent processes defined by the system, in the topological space.

The actions *out (process, resource, lock/unlock)* that a certain process performs correspond to the set of coordinate points on each axis composed by an ordered sequence of real numbers between 0 and 1, representing the scheduled actions (a transaction) that a given process will execute.

For every concurrent process in the MSC, each action *out (process, resource, lock)* upon the shared instance *resource*, will be identified with a lock (P) action and each action *out (process, resource, unlock)* upon the shared instance *resource*, will be identified with an unlock (V) action. Consequently for each concurrent process, there will be its corresponding partially ordered actions, into a sequence of ordered real numbers along the axis interval [0,..,1]. The resources identified in each action *out (process, resource, lock/unlock)* will be labeled shared resources that are being locked and unlocked.

Once this mapping is realized, the progress graph is created. The next step is the application of the deadlock detection algorithm [15] to each resulting topological spaces. Figure 7 is a simple illustration of the correspondence between MSC and progress graph, with two scenarios identified by the topological technique: a safe execution path (1) and a deadlock situation (2).

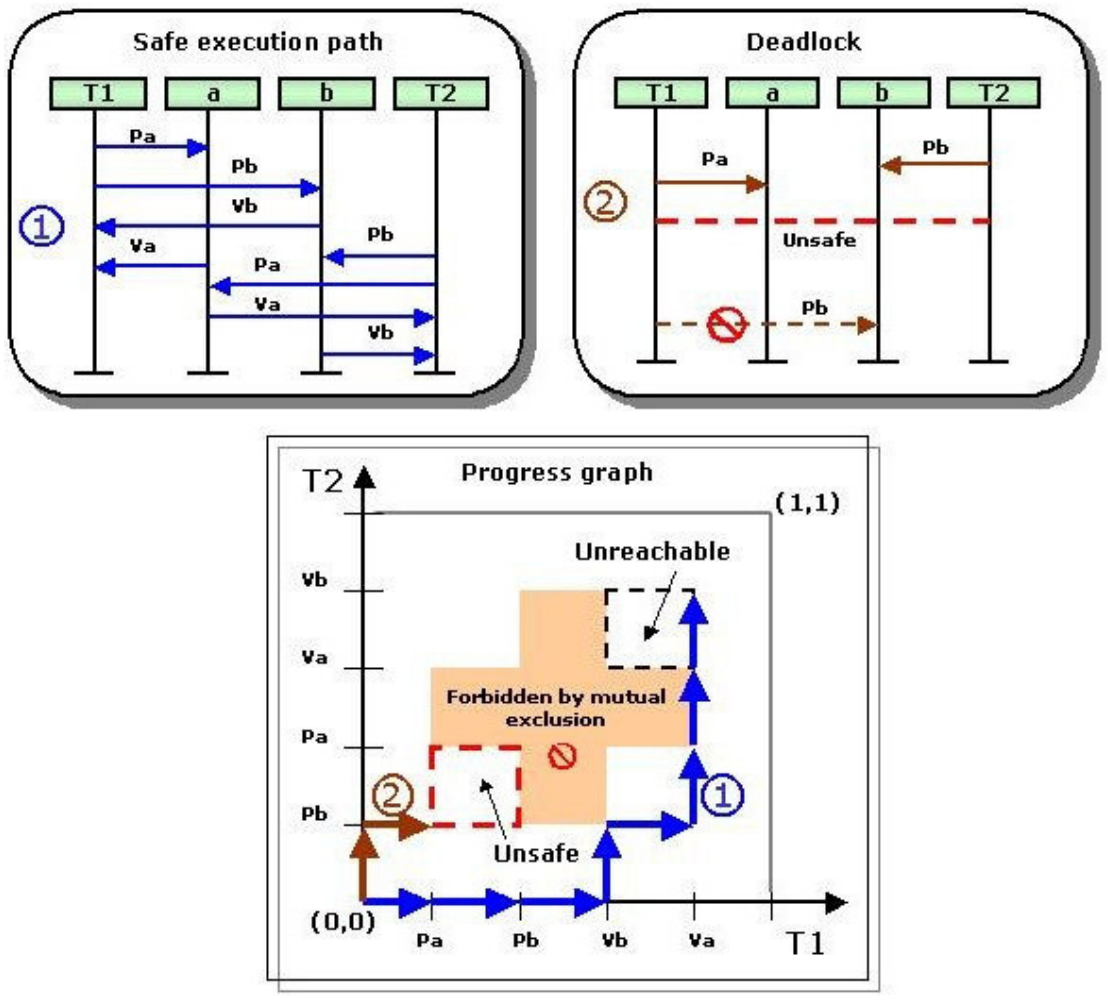

Figure 7: Correspondence between MSC and progress graph

## 5. CONCLUSIONS AND FUTURE PROSPECTS

This paper proposes a simple "proof-of-concepts" prototype to formally treat concurrency in real time systems by considering the integration of a deadlock detection algorithm based on topology, to MSC. One of the initial results of this integration is the possibility of formally verifying MSC scenarios and reliably finding forbidden scenarios at the early phases of development.

The use of geometric and algebraic topology concepts allows to promptly identifying critical regions for decision making considerations. The use of MSC as a language to express scenarios of a system practically provides a one-to-one correspondence between both formalisms. This fact guarantees that the use of the proposed method does not affect the necessary flexibility. Reliability is also accomplished by the implementation of a simple and precise algorithm derived from the geometrical configuration of the state space of the system.

The integration of formal methods with behavioral models that are flexible enough to be used at the early phases of the software development is a step forward in the characterization of a rigorous treatment of concurrency. Although the method was proposed to be primarily applied during the system requirements analysis, it can also be used to formally verify refined MSC scenarios in the detailed design.

There are still many aspects to be considered in this research. Future implementation of the prototype has to aim at the development of a friendly user interface. The proposed method can also be extended with apparently minor modifications in order to implement the case of semaphores. The use of more advanced topological techniques, such as "direct homotopy", to more generalized and/or complex situations, like those involving a varying number of concurrent processes, is a promising line of research in the emergent discipline of topological concurrency theory.